\documentclass[journal=apchd5,manuscript=letter]{achemso}

\usepackage[version=3]{mhchem}%
\author{Skender Morina}
\affiliation[University of Iceland]{Science Institute,
University of Iceland, Dunhagi 3, IS-107, Reykjavik, Iceland}
\alsoaffiliation[ITMO University]{ITMO University, 
Kronverkskiy prospekt 49, Saint Petersburg 197101, Russia}

\email{skender@hi.is}
\author{Kevin Dini}
\affiliation[University of Iceland]{Science Institute,
University of Iceland, Dunhagi 3, IS-107, Reykjavik, Iceland}
\alsoaffiliation[ITMO University]{ITMO University, Kronverkskiy prospekt 49, Saint Petersburg 197101, Russia}

\author{Ivan V. Iorsh}
\affiliation[ITMO University]{ITMO University, Kronverkskiy prospekt 49, Saint Petersburg 197101, Russia}
\author{Ivan A. Shelykh}
\affiliation[University of Iceland]{Science Institute,
University of Iceland, Dunhagi 3, IS-107, Reykjavik, Iceland}
\alsoaffiliation[ITMO University]{ITMO University, Kronverkskiy prospekt 49, Saint Petersburg 197101, Russia}

\title[Optical trapping of electrons in graphene]
  {Optical trapping of electrons in graphene}

\keywords{Graphene, electron transport, optical trapping}

\begin{document}

\begin{abstract}
We propose an experimentally friendly scheme for trapping quasi- relativistic electrons in graphene by an electromagnetic beam with circular polarization and spatially inhomogeneous profile with an intensity dip. The trapping is achieved due to the effect of bandgap opening outside the trapping region. The proposed mechanism allows for non- invasive electron confinement in graphene without any need of the chemical patterning of the sample or the application of metallic gates. 
\end{abstract}

\section{Introduction}
Since the experimental discovery of graphene, its unique electronic properties continue to attract an increasing interest of
the scientific community \cite{Geim_07}. The spectrum of the electronic states around \textit{K} and \textit{K$^{\prime}$} points in the Brillouin zone consists of a pair of touching Dirac cones, and thus mimics the dispersion of massless relativistic Fermions. This fact has dramatic consequences on the transport properties of graphene, one of which is Klein tunneling: the perfect transmission of gapless Dirac electrons through arbitrary potential barriers at normal incidence\cite{Klein,Dombey}. 

Being extremely interesting from the point of view of fundamental physics, Klein tunneling nevertheless poses serious problems for a wide range of practical applications of graphene where confinement of electrons is necessary. To circumvent this obstacle, a variety of methods has been proposed. They include chemical functionalization \cite{Withers,Balog}, mechanical cutting of the graphene sheet into nanoribbons or nanodisks \cite{Brey} and application of local strain resulting in the onset of an artificial gauge field \cite{Ni,Cocco}. Most of those methods, however, need irreversible modification of the graphene sheet and do not allow for a controllable tuning of the trapping parameters such as the strength of confinement. 

In the present paper we explore an alternative way to achieve the  trapping of massless Dirac electrons using fully optical means. Optical trapping is a standard way of the preparation of cold atom lattices (see e.g. Refs.~\cite{Morsch2006, Dholakia2010} for the review) and the confinement of nanoparticles by optical tweezers (see e.g. Ref.~\cite{Marago2013} and references therein). In the domain of condensed matter, the basis for the optical trapping is provided by the possibility to modify the energy spectrum of a material system by strong coupling to the high- frequency laser radiation resulting in the  dynamic Stark effect. Dramatic modifications of the transport properties in the regime of strong light- matter coupling were reported for semiconductor quantum wells 
\cite{Wagner_10,Wagner_13,Morina_15,Dini_16,Pervishko_15}, carbon nanostructures 
\cite{Monroy_16,Kibis_16,Kristinsson_16,Iurov_13,Ezawa_13,Iorsh_17,Kibis_17}, topological 
insulators \cite{Usaj_14,Torres_14,Calvo_15,Yudin_16} and others.

In particular, it was shown that in graphene strongly coupled to circular polarized light the bandgap $\Delta_g$ opens \cite{Oka2009,Kibis2010,UsajII_14}. Its value depends on the intensity $I$ and frequency $\omega$ of the driving field and can thus be reversibly changed in a controllable way. In realistic configurations this light induced bandgap can reach several meV. 

\begin{figure}
\includegraphics[width=\columnwidth]{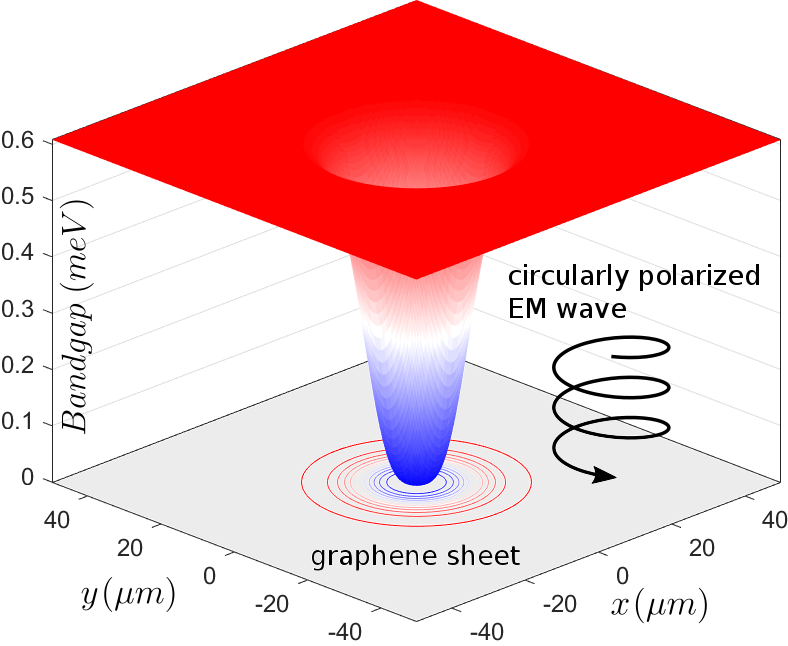} 
\caption{The profile of the bandgap in real space 
induced by an inhomogeneous high-frequency electromagnetic field with the intensity having a Guassian dip. The value of the optically induced gap is proportional to the intensity of the dressing field, so  electrons become trapped in the region where intensity is minimal.}
\label{Fig1}
\end{figure}

This effect can be exploited for the trapping of Dirac electrons in graphene. Indeed, consider the case when the intensity of the driving field is not homogeneous and has a minimum in real space at $r=0$. In this situation the effective bandgap induced by light- matter coupling will be minimal in the center of the dip and increase if one deviates further from it as is shown in the Fig.~\ref{Fig1}. This will confine the low energy electrons in the region around $r=0$. The mechanism of the confinement is similar to that obtained in semiconductor heterostructures when a layer of narrow band semiconductor is sandwiched between wide gap semiconductors with the only difference being that the band mismatch in our case is produced  all optically.  

\section{The model}

Let us consider a monolayer of graphene, which lies in the plane $\mathbf r = (x,y)$ at $z=0$ and interacts with an electromagnetic wave propagating along the z-axis.  The frequency of the wave $\omega$ is assumed to be high enough to satisfy the condition $\omega\tau\gg1$, where $\tau$ is a characteristic relaxation time in the system. In this case, the electromagnetic wave can not be absorbed around the Dirac points and should be considered as a pure dressing field. 

Let us consider a monolayer of graphene, which lies in the plane $\mathbf r = (x,y)$ at $z=0$ and interacts with an electromagnetic wave propagating along the z-axis.  The frequency of the wave $\omega$ is assumed to be high enough to satisfy the condition $\omega\tau\gg1$, where $\tau$ is a characteristic relaxation time in the system. In this case, the electromagnetic wave can not be absorbed around the Dirac points and should be considered as a pure dressing field. The low- energy Hamiltonian of the system reads: 
\begin{equation}\label{H0}
\hat{{\cal H}}(t)=\hbar v_F\left[\xi \sigma_x \left(k_x+\frac{eA_x(t)}{\hbar}\right) + \sigma_y \left(k_y+\frac{eA_y(t)}{\hbar}\right)\right] 
\end{equation}
where $v_F$ is the Fermi velocity, $\xi$ the valley index and $\sigma_i$, $i = x,y$, are the Pauli matrices and $\xi=\pm1$ is a valley factor. The interaction with external electromagnetic radiation was introduced via the minimal coupling substitution,  
$k_{x,y} \rightarrow k_{x,y} + (e/\hbar) A_{x,y}  $ where $A_{x,y}$  corresponds to the vector potential of the dressing field.  To break time 
reversal symmetry and open the band gap at the Dirac points we should consider the case of circular polarization, choosing
\begin{align}\label{VectorPotential}
A_x=\frac{c E_0}{\omega}G\left(r\right)\sin\left(\omega t \right),\\
A_y=-\frac{c E_0}{\omega}G\left(r\right)\cos\left(\omega t \right),
\end{align}
where $\omega$ and $E_0$ are  frequency and amplitude of the dressing field, and function $G(r)$ describes its profile in the real space. To be specific, we choose the latter to be represented by  a Gaussian dip,
\begin{equation}\label{GeometryTrap}
G\left(r\right) = 1- \exp\left(-\frac{r^2}{2 L^2}\right)
\end{equation}
where parameter $L$ characterizes the lateral size of the intensity dip. 

The Hamiltonian in Eq.~\ref{H0} is time-dependent, but in the high frequency limit it can be reduced to a stationary effective Hamiltonian. The mathematical basis for that is provided by Floquet theory of periodically driven quantum systems 
\cite{Hanggi_98,Kohler_05,Bukov_15,Holthaus_16}. 
The main steps are as follows. The time-dependent Hamiltonian can be 
expressed as : 
\begin{equation}\label{SeparationHarmonics}
\hat{\cal H}\left(r,t\right)= \hat{ \cal H}_0+ 
\hat{V} \exp\left(i \omega t \right)+ 
\hat{V}^\dagger \exp\left(-i \omega t \right)
\end{equation}
where
\begin{eqnarray}\label{Harmonics}
\hat{{\cal H}}_0(t)=\hbar v_F\left[\xi \sigma_x k_x+ \sigma_y k_y\right],\\
\hat{V}\left(r\right)= \frac{\hbar\Omega}{2}\left( \xi \sigma_x - 
i \sigma_y\right)  G\left(r\right).
\end{eqnarray}
and
\begin{equation}\label{Dimensionless}
\hbar\Omega = \frac{v_F e E_0}{\omega}
\end{equation}
is a parameter describing the strength of electron- photon coupling which can be referred to as a characteristic Rabi energy. Since the frequency $\omega$ is assumed to be high compared to all characteristic frequencies of the system, the electron dynamics is not able to follow the fast time oscillations of the vector potential, and the effective time-independent Hamiltonian can be obtained by Floquet-Magnus expansion
\cite{Casas_01,Goldman_14,Eckardt_15} in powers of $\omega^{-1}$. Restricting ourselves to the first three terms in the infinite series we get:
\begin{eqnarray}
\nonumber\hat{{\cal H}}_{\mathrm{eff}}\approx\hat{{\cal H}}_0+\frac{\left[\hat{V},
\hat{V}^\dagger\right]}{\hbar\omega}+
\frac{ \left[\left[\hat{V},\hat{{\cal H}}_0\right],\hat{V}^\dagger\right]+
\mathrm{H.c.}}{2(\hbar \omega)^2}\approx\\ \nonumber
\approx \hbar \widetilde{v}_F(r)( \xi \sigma_x k_x + 
\sigma_y k_y )  - \xi \frac{\Delta_g(r)}{2} \sigma_z \\
 + i\frac{\hbar v_F\Omega^2}{\omega^2L^2}
\left(\xi\sigma_xx + \sigma_yy \right)\exp\left(-\frac{r^2}{L^2}\right),
\label{Effective}
\end{eqnarray}
In this equation $\Delta_g(r) = 2\hbar\Omega^2 G^2(r)/\omega$ is the position- dependent gap, which is a monotonously increasing function of $r$ with $\Delta(0)=0$ and 
\begin{equation}
\Delta(\infty)=\frac{2\hbar\Omega^2}{\omega}=\frac{4v_F^2e^2I}{\hbar\epsilon_0c\omega^3}
\label{Difty}
\end{equation}
where $I=\epsilon_0 c E_0^2/2$ is the intensity of the dressing field. The corresponding term the Eq.~\ref{Effective} is responsible for the trapping of particles with energies $E<\Delta(\infty)$. The position- dependent renormalized Fermi velocity $\widetilde{v}_F(r)=v_F(1-\Omega^2\omega^{-2} G^2(r))$. The last term in Eq.~\ref{Effective} appears due to the energy- momentum non-commutativity. It is small with respect the other terms but should be nevertheless retained in order to keep the effective Hamiltonian Hermitian.

To make the trapping most efficient, one would wish to produce deep traps with small lateral size, minimizing $L$ and maximizing $\Delta(\infty)$. Unfortunately, these two parameters are not completely independent. Indeed, according to Eq.~\ref{Difty}, the value of the gap equal to the depth of the trap is inversely proportional to the cube of the frequency of the dressing field. Therefore, if one wants to keep the intensity $I$ moderate one can not use very high frequencies. On the other hand, for a given frequency the lateral size of the trap can not be done arbitrary small because of the diffraction limit and its minimal size can be estimated as 
\begin{equation}
L_{min}\approx\lambda=\frac{2\pi c}{\omega}
\end{equation}
In principle, this size can be further reduced by using the methods of subwavelength optics, but this will need metallic patterning of the sample and consideration of this case goes beyond the scope of the present work.
Our estimations show that there is an optimal range of frequencies corresponding to THz and far infrared for which the traps with a depth of one to several meV with lateral size of tens of microns can be achieved for realistic values of the dressing intensities not exceeding several kW/cm$^2$.

\begin{figure*}
\includegraphics[width=\textwidth]{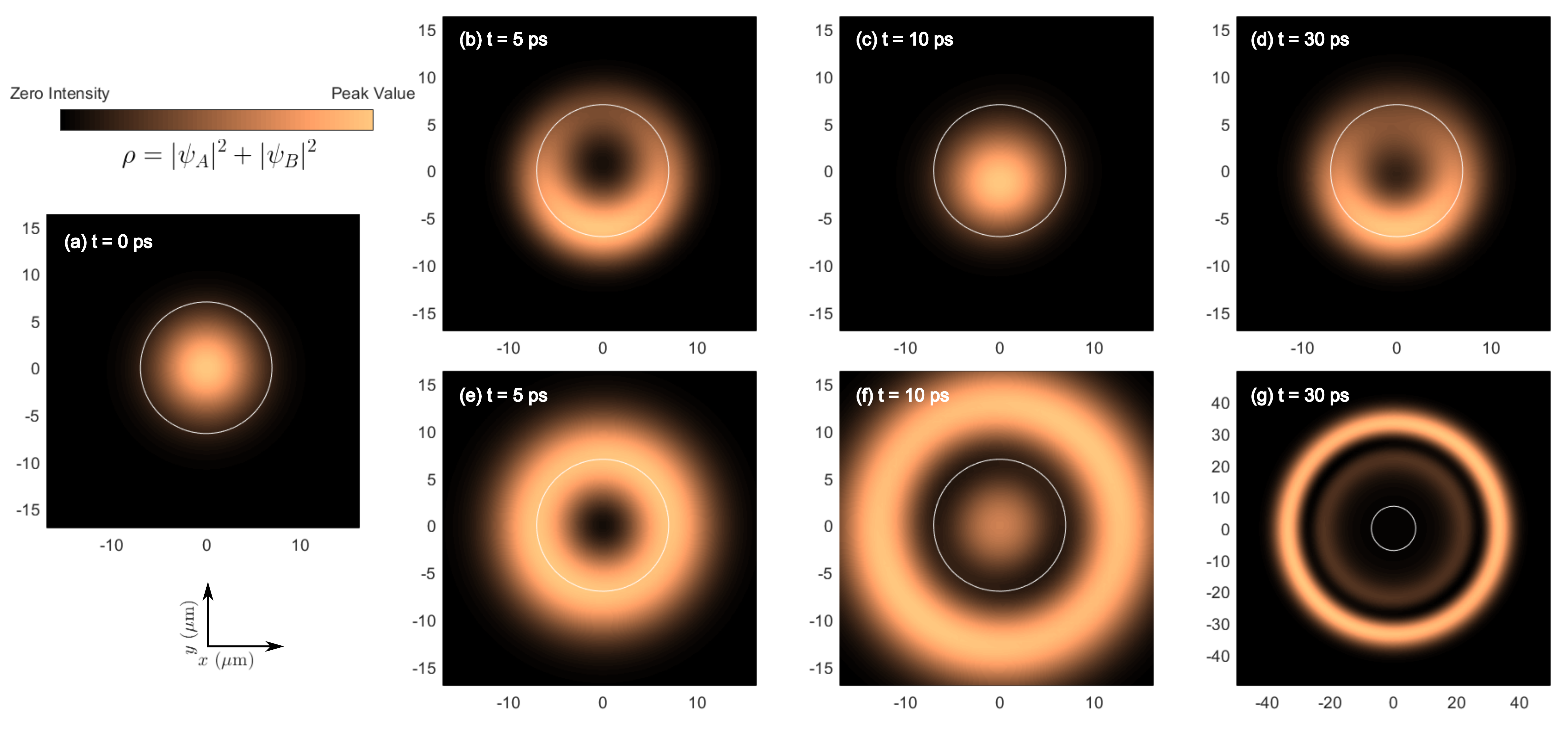}
\caption{Plots of the total electron density, $\rho=|\psi_A|^2 + |\psi_B|^2$, for several values of the evolution time
indicated in each subfigure. Panel \textbf{(a)} corresponds to the initial distribution at $t=0$. Panels \textbf{(b)-(d)} correspond to the dynamics in the ðpresence of the optical trap. Panels 
 \textbf{(e)-(g)} correspond to the free electron propagation (note that panel (g) includes a larger area of the system). One clearly sees that the presence of the optical trap effectively confines the electronic wavepacket which stays localized around $r=0$ at all times. }
\label{fig:Intensity}
\end{figure*}

\section{Results and discussion}
To demonstrate the effectiveness of the proposed trapping scheme, we study numerically the dynamics of the electronic wavepacket initially located around $r=0$. We use the following parameters of the dressing field; intensity $I =$~330~W/cm$^2$, dressing field frequency $\omega =15$~THz, and intensity dip radius $L = 7~\mu$m. The width of the initial wavepacket was taken to be $d = 5~\mu$m so that it has energy distribution of $\hbar v_F/d \approx 0.13$~meV around the Fermi energy $E_F = 0$~meV. The initial wavepacket is composed of equal contributions from the two equivalent valleys ($K$ and $K'$) in the band structure of graphene.
The results are shown in the Fig.~\ref{fig:Intensity}.  As it is clear from the panels (b)-(d), for the case of the electromagnetic dressing the electrons stay around $r=0$ where the gap is smaller and can not penetrate to the region where the gap reaches its maximal value. Panels (f)-(g) correspond to the case of the freely propagaing wavepacket and are demonstrated for the reasons of comparison. Wavepacket dynamics in this case reproduces well known results studied elsewhere before \cite{Maksimova08}.
We also estimated the position of the energy levels in our trapping potential as it is shown at the Table~ \ref{tab:resonances}. Characteristic separation between the neighboring levels is of the order of magnitude of 0.1 meV which should be possible to observe experimentally.
\begin{table}[H]
\caption{The energies of the lowest confined states for the optical trap in graphene. See main text for the values of the parameters. The energy levels are found by assuming a perfect trap and imposing vanishing boundary conditions on the wavefunction at the radius of the trap.}
\begin{tabular}{c c c c}
$j$ & $k_j (\mu\textrm{m}^{-1})$ & $E_j$ (meV) & $E_{j} - E_{j-1}$ (meV)  \\ \hline
1 & 0.347 & 0.188 &\\ 
2 & 0.547 & 0.299 & 0.111\\ 
3 & 0.734 & 0.401 & 0.102\\ 
4 & 0.789 & 0.430 & 0.030\\ 
5 & 1.002 & 0.547 & 0.117\\ \hline
\end{tabular}
\label{tab:resonances}
\end{table}

\begin{acknowledgement}

This work was supported by the megagrant 14.Y26.31.0015 and Goszadanie no. 3.2614.2017/4.6 of the Ministry of Education and Science of Russian Federation,  Icelandic Research Fund, Grant No. 163082-051 and Horizon2020 project CoExAN. K.D. thanks National University of Belarus for hospitality during the work on the project.
\end{acknowledgement}

\begin{suppinfo}

\end{suppinfo}


\begin{thebibliography}{99}

\bibitem{Geim_07} A. K. Geim and K. S. Novoselov, The rise of graphene, Nature Mater. 
\textbf{6}, 183 (2007).

\bibitem{Klein}
O. Klein, Die Reflexion von Elektronen an einem Potentialsprung nach der 
relativistischen Dynamik von Dirac, Z. Phys \textbf{53}, 157-165 (1929).

\bibitem{Dombey}
N.Dombey, A.Calogeracos, Seventy years of the Klein paradox. 
Phys. Rep. \textbf{315}, 41-58 (1999).

\bibitem{Withers}
F. Withers, M. Dubois, and A.K. Savchenko, Electron properties of fluorinated 
single-layer graphene transistors, Phys. Rev. B \textbf{82} 073403 (2010). 

\bibitem{Balog}
R. Balog \& al., Bandgap opening in graphene induced by patterned hydrogen 
adsorption, Nat. Mater. \textbf{9}, 315–319 (2010).

\bibitem{Brey}
L. Brey and H. A. Fertig, Electronic states of graphene nanoribbons studied 
with the Dirac equation, Phys. Rev. B \textbf{73}, 235411 (2006)

\bibitem{Ni}
Z.H. Ni, T. Yu, Y.H. Lu, Y.Y. Wang, Y.P. Feng, and Z.X. Shen, Uniaxial Strain 
on Graphene: Raman Spectroscopy Study and Band-Gap Opening, 
ACS Nano \textbf{2}, 2301 (2008).   

\bibitem{Cocco}
G. Cocco, E. Cadelano, and L. Colombo, Gap opening in graphene by shear strain,
 Phys. Rev. B \textbf{81}, 241412(R) (2010).

\bibitem{Morsch2006}
Oliver Morsch and Markus Oberthaler, Rev. Mod. Phys. \textbf{78}, 179 (2006)

\bibitem{Dholakia2010}
Kishan Dholakia and Pavel Zemanek, Rev. Mod. Phys. \textbf{82}, 1767 (2010)

\bibitem{Marago2013}
O. M. Marago, P. H. Jones, P. G. Gucciardi, G. Volpe, and A. C.
Ferrari, Nat. Nanotechnol. \textbf{8}, 807 (2013).

\bibitem{Wagner_10}
M.Wagner, H. Schneider, D. Stehr, S.Winnerl, A.M. Andrews,S. Schartner, 
G. Strasser, and M. Helm, Observation of the Intraexciton Autler-Townes Effect 
in GaAs/AlGaAs Semiconductor Quantum Wells, 
Phys. Rev. Lett. \textbf{105}, 167401 (2010).

\bibitem{Wagner_13}
M. Teich, M. Wagner, H. Schneider, and M. Helm, Semiconductor quantum well 
excitons in strong, narrowband terahertz fields, 
New J. Phys. \textbf{15}, 065007 (2013).

\bibitem{Morina_15}
S. Morina, O. V. Kibis, A. A. Pervishko, and I. A. Shelykh, Transport properties
 of a two-dimensional electron gas dressed by light, 
 Phys. Rev. B \textbf{91}, 155312 (2015).

\bibitem{Pervishko_15}
A. A. Pervishko, O. V. Kibis, S. Morina, and I. A. Shelykh, Control of spin 
dynamics in a two-dimensional electron gas by electromagnetic dressing, 
Phys. Rev. B \textbf{92}, 205403 (2015).

\bibitem{Dini_16}
K. Dini, O. V. Kibis, and I. A. Shelykh, Magnetic properties of a 
two-dimensional electron gas strongly coupled to light, 
Phys. Rev. B \textbf{93}, 235411 (2016).

\bibitem{Monroy_16}
R. Vega Monroy and G. Salazar Cohen, Photon-Induced Quantum Oscillations of the 
Terahertz Conductivity in Graphene, Nano Letters \textbf{16} (11), 6797 (2016).

\bibitem{Kibis_16}
O. V. Kibis, S. Morina, K. Dini, and I. A. Shelykh, Magnetoelectronic properties
 of graphene dressed by a high-frequency field, Phys. Rev. B 93, 115420 (2016).

\bibitem{Kristinsson_16}
K. Kristinsson, O.V. Kibis, S. Morina, and I. A. Shelykh, Control of electronic
 transport in graphene by electromagnetic dressing, Sci. Rep. \textbf{6}, 20082
 (2016).

\bibitem{Kibis_17}
 O. V. Kibis, K. Dini, I. V. Iorsh and I. A. Shelykh, All-optical band engineering of gapped Dirac materials, Phys. Rev. B, \textbf{95} 125401 (2017).
 
\bibitem{Iorsh_17}
I. V. Iorsh, K. Dini, O. V. Kibis, and I. A. Shelykh, Optically induced Lifshitz transition in bilayer graphene, Phys. Rev. B 96, 155432 (2017).
 
\bibitem{Iurov_13}
 A. Iurov, G. Gumbs, O. Roslyak and D. Huang, Photon dressed electronic states in topological insulators: tunneling and conductance, J. Phys.: Condensed Matter \textbf{25}, 135502 (2013).
 
\bibitem{Ezawa_13}
M. Ezawa, Photoinduced Topological Phase Transition and a Single Dirac-Cone State in Silicene, Phys. Rev. Lett. \textbf{110}, 026603 (2013).

\bibitem{Usaj_14}
G. Usaj, P. M. Perez-Piskunow, L. E. F. Foa Torres, andC. A. Balseiro, 
Irradiated graphene as a tunable Floquet topological insulator, 
Phys. Rev. B \textbf{90}, 115423 (2014).

\bibitem{Torres_14}
L. E. F. Foa Torres, P. M. Perez-Piskunow, C. A. Balseiro, and G. Usaj, 
Multiterminal Conductance of a Floquet Topological Insulator, 
Phys. Rev. Lett. \textbf{113}, 266801 (2014).

\bibitem{Calvo_15}
H. L. Calvo, L. E. F. Foa Torres, P. M. Perez-Piskunow,C. A. Balseiro, and 
G. Usaj, Floquet interface states in illuminated three-dimensional topological 
insulators, Phys. Rev. B \textbf{91}, 241404(R) (2015).

\bibitem{Yudin_16}
D. Yudin, O. V. Kibis, and I. A. Shelykh, Optically tunable spin transport on 
the surface of a topological insulator, New J. Phys. \textbf{18}, 103014 (2016).

\bibitem{Oka2009} T. Oka and H. Aoki, Phys. Rev. B 79, 081406(R) (2009).

\bibitem{Kibis2010} O. V. Kibis, Phys. Rev. B 81, 165433 (2010).

\bibitem{UsajII_14}  G. Usaj, P. M. Perez-Piskunow, L. E. F. Foa Torres, and
C. A. Balseiro, Irradiated graphene as a tunable Floquet topological insulator, Phys. Rev. B.
\textbf{90}, 115423 (2014).

\bibitem{Hanggi_98}
P. H\"anngi, Driven quantum systems, in {\it Quantum Transport and
Dissipation} edited by T. Dittrich, P. H\"anggi, G.-L. Ingold, B.
Kramer, G. Sch\"on, and W. Zwerger (Wiley, Weinheim, 1998).

\bibitem{Kohler_05}
S. Kohler, J. Lehmann, and P. H\"anngi, Driven quantum transport
on the nanoscale, Phys. Rep. {\bf 406}, 379--446 (2005).

\bibitem{Bukov_15} M. Bukov, L. D'Alessio, and A. Polkovnikov, Universal 
High-Frequency Behavior of Periodically Driven Systems:
from Dynamical Stabilization to Floquet Engineering, Adv. Phys.
\textbf{64}, 139--226 (2015).

\bibitem{Holthaus_16}
M. Holthaus, Floquet engineering with quasienergy bands of
periodically driven optical lattices, J. Phys. B {\bf 49}, 013001
(2016).
 
\bibitem{Casas_01}
F. Casas, J. A. Oteo, and J. Ros, Floquet theory: exponential perturbative 
treatment, J. Phys. A \textbf{34}, 16 (2001).

\bibitem{Goldman_14}
N. GoldMan and J. Dalibard, Periodically Driven Quantum Systems: Effective 
Hamiltonians and Engineered Gauge Fields, 
Phys. Rev. X \textbf{4}, 031027 (2014).

\bibitem{Eckardt_15}
A. Eckardt and E. Anisimovas, High-frequency approximation for periodically 
driven quantum systems from a Floquet-space perspective, 
New J. Phys. \textbf{17}, 093039 (2015).

\bibitem{Maksimova08} G. M. Maksimova, V. Ya. Demikhovskii, and E. V. Frolova,
Wave packet dynamics in a monolayer graphene, Phys. Rev. B \textbf{78},
235321 (2008).

\end{thebibliography}
\end{document}